\documentclass[12pt]{iopart}
\usepackage[pdftex]{graphicx}
\usepackage{epstopdf}
\graphicspath{{./image/}}

\usepackage{array}
\usepackage{rotating}
\usepackage{url}

\usepackage{multirow}

\usepackage{multicol}

\usepackage{soul}

\usepackage{lscape}

\usepackage[normalem]{ulem}
\usepackage{iopams}
\expandafter\let\csname equation*\endcsname\relax
\expandafter\let\csname endequation*\endcsname\relax
\usepackage[cmex10]{amsmath}
\usepackage{amssymb}

\usepackage{makecell}
\usepackage{textcomp}

\usepackage{lipsum}

\usepackage{capt-of}

\usepackage{tabularx}
\usepackage{bm}
\usepackage{calc}

\usepackage{xcolor,colortbl}

\usepackage{sidecap, caption}
\usepackage{tabu}
\usepackage{mdframed}

\usepackage{lscape}
\usepackage{cite}

\usepackage{rotating}

\usepackage{tikz}
\usetikzlibrary{shapes.geometric}
\usetikzlibrary{calc}
\newcommand{\toppage}{
	\begin{tikzpicture}[remember picture,overlay]
	\node[align=left, anchor=north west]
	at ($(current page.north west) + (2,-1)$)
	{\texttt{This article has been published in the \textit{Physiological Measurement} journal}};
	\end{tikzpicture}
}

\begin{document}
	
	\title[Automatic Sleep Staging of EEG Signals]{Automatic Sleep Staging of EEG Signals: Recent Development, Challenges, and Future Directions}
	
	\author{Huy Phan$^{1,2}$, Kaare Mikkelsen$^{3}$}
	
	\address{$^1$School of Electronic Engineering \& Computer Science, Queen Mary University of London, UK}
	\address{$^2$The Alan Turing Institute, UK}
	\address{$^3$Department of Electrical and Computer Engineering, Aarhus University, Denmark}% \\
	\ead{h.phan@qmul.ac.uk, mikkelsen.kaare@ece.au.dk}
	\vspace{10pt}
	
	\begin{abstract} 
		
		Modern deep learning holds a great potential to transform clinical practice on human sleep. Teaching a machine to carry out routine tasks would be a  tremendous reduction in workload for clinicians. Sleep staging, a fundamental step in sleep practice, is a suitable task for this and will be the focus in this article. Recently, automatic sleep staging systems have been trained to mimic manual scoring, leading to similar performance  to human sleep experts, at least on scoring of healthy subjects. Despite  tremendous progress, we have not seen automatic sleep scoring adopted widely in clinical environments. This review aims to give a shared view of the authors on the most recent state-of-the-art development in automatic sleep staging, the challenges that still need to be addressed, and the future directions for automatic sleep scoring to achieve clinical value. 
		\toppage
	\end{abstract}
	
	\section{Introduction}
	
	Sleep makes up almost one third of our lives. Good sleep is crucial in maintaining one's mental and physical health \cite{Siegel2005,Maquet2001} while sleep disorders are linked with a host of different ailments \cite{perez-pozuelo_future_2020}. Screening, assessment, and diagnosis for sleep disorders require each 30-second epoch of an overnight polysomnogram (PSG) to be assigned a sleep stage. This procedure is known as sleep staging or scoring. The sequence of sleep stages is critical for measuring parameters of the sleep macrostructure, such as the sleep cycles, the time spend in each stage, sleep latency, wake after sleep onset (WASO), etc. Sleep stages and cycles, which manifest the underlying neuro-physiological processes, are also a rich source for mining diagnostic markers for a wide range of sleep disorders \cite{Norman2006, Cooray2019, Stephansen2018, Christensen2015}, from a common one like obstructive sleep apnea (OSA) \cite{Senaratna2017,Redline1999} to a rare one like narcolepsy \cite{Stephansen2018, Christensen2015}.
	
	Sleep staging is still largely carried out by clinicians in sleep clinics guided by a well established manual, published by the American Academy of Sleep Medicine \cite{berry_aasm_2016}. This labor-intensive and time-consuming manual scoring is unsuited for handling large-scale data and cannot be scaled to serve the needs of millions suffering from sleep disorders \cite{Krieger2017,Chattu2019,InstMed2006}. At the same time, there is an increasing need for longitudinal monitoring in home environments. Accurate and cost effective monitoring of sleep not only has great medical value but also allows individuals to self-assess and self-manage their sleep. Thus, it is imperative for sleep staging to be automated. The fact that it follows a predefined set of rules makes sleep staging a perfect task for automation with machine learning. Furthermore, a machine can perform the task thousands of times faster than a human expert, while saving a clinician thousands of hours a year and making sleep assessment and diagnosis more widely available.
	
	Indeed, given the standardization of data through the use of PSG for recordings, a very large volume of methods development has already taken place. Particularly, the existence of increasingly large public data sets online (for example, PhysioNet \cite{Goldberger2000} and the National Sleep Research Resource (NSRR) \cite{Zhang2018}) has enabled exploitation of deep learning \cite{LeCun2015,Goodfellow2016} to teach a machine to perform sleep staging using a large amount of training data in the last couple of years. These efforts have led to more advanced and sensible methods \cite{Supratak2017, Phan2019a, Biswal2018a, Olesen2021, guillot_robustsleepnet_2021, Phan2021a, Phan2021b} which have surpassed the agreement level of experts' scoring and achieved a performance acceptable for clinical use. Notwithstanding this tremendous progress, machine sleep scoring still needs to overcome several technical and clinical barriers to be widely adopted and deliver full clinical value. We see great potential for further development. First, from algorithmic perspectives, we consider sleep staging to be an interesting modelling problem where novel methods can be developed to tackle the foreseeable obstacles and pave the way for clinical usage. Second, this is related to the emergence of new recording platforms \cite{miettinen_success_2018,arnal_dreem_2019,mikkelsen_accurate_2019,mikkelsen_machine-learning-derived_2018} to simplify the sleep setup for home-based monitoring purposes. 
	
	In this review article, we begin by discussing the clinical context of automatic sleep scoring, after which we  give an overview as well as technical insights of the state-of-the-art methods for automatic scoring of EEG data. Readers should note that a few existing reviews, such as Fiorilli \emph{et al.} \cite{Fiorillo2019} and Faust \emph{et al.} 2019 \cite{Faust2019}, have summed up the topic prior to 2019. A review on broader applications of deep learning on EEG analysis also exists \cite{Roy2019}. To avoid re-inventing the wheel, this article focuses on the latest method development in automatic sleep staging. In addition, we limit the scope of this article to fine-grained (\emph{i.e.}, five stages) sleep staging using PSG and modalities directly reading brain activities, such as mobile EEGs, and will not cover research work using other modalities, such as electrocardiogram (ECG)/photoplethysmography (PPG) \cite{Radha2021}, actigraphy \cite{Zhai2020}, audio \cite{Dafna2018}, video \cite{Long2019}, and radar \cite{Toften2020,Piriyajitakonkij2021}. We then discuss the current challenges, and suggest directions to move forwards. As we shall discuss in the next section, much good work has already been done on this problem. However, as it will become clear in the rest of this review, we believe the field has only solved the first, ``entry'', problem, and a plethora of new and exciting tasks lie ahead of us.
		
	\section{Clinical context}
	
		Manual sleep scoring is a somewhat reliable, highly versatile method, which readily yields interpretations and which is standardized across the world. This has made it a good solution, but also a local optimum which is hard to escape. By this we mean that it is not the best possible solution, due to a number of drawbacks: 
		
		\begin{enumerate}
			\item It is very time consuming  (and therefore expensive) to manually score an entire night's recording. Even more so if sleep events are also to be annotated.
			\item Despite the existence of a sleep scoring standard, there is still variation between individual scorers.
			\item The sleep scoring manual is based on the polysomnography (PSG) recording setup, which is generally considered to be unwieldy and invasive on sleep. This, combined with the cost of each recording, means that clinicians will usually have to `make do' with a single (at most two) nights of data, which may not be as representative of the patient's usual nights as one would hope.
		\end{enumerate}
		
		It should come as no surprise that the properties of manual sleep scoring have shaped the way sleep monitoring is used - few recordings per subject, qualitative (non-data driven) analysis. This can make it hard, within the clinical reality, to immediately see the benefits of a new method (automatic sleep scoring with other sensor setups). Figuratively speaking, if you have learned to solve all problems using nails, it is hard to see how a screwdriver can compete with your hammer.
		
		Automatic sleep scoring can reduce the costs of existing procedures (PSG recordings either in lab or at home), but also open the door to new ways of using sleep clinically, which today would be infeasible. We can imagine population wide screening for early stages of debilitating diseases (e.g., the REM Sleep Behavior Disorder (RBD) is known to be tightly associated with Parkinson's disease \cite{lin_rbd_2018}), or routine follow-up procedures quantifying patient sleep after they leave the hospital. These procedures could have very real clinical benefits, but they all require changes to how sleep recordings are used and managed, not to abolish existing procedures, but to supplement them. 
		
		An algorithm-first approach to clinical sleep can also solve other problems. First, given the costs of a PSG recording, clinicians may often have to `make do' with whatever recordings they get, even if the quality is questionable. However, if the standard quantum becomes a week's worth of data, automatic discarding of low quality nights would be trivial. Second, definitions surrounding sleep have been developed and evaluated based on how well they can be used in manual sleep scoring. Computers have far fewer restrictions in this manner, and we can imagine more flexible taxonomies, such as hypnodensity plots \cite{Stephansen2018} or even disease-specific sleep states. 
	
	\section{The state-of-the-art sleep scoring}
	
	Modern deep learning \cite{LeCun2015,Goodfellow2016} crept into sleep research much slower than other fields such as computer vision, natural language processing, and speech recognition. The use of deep neural networks for automatic sleep staging only started around 5 years ago even though their resurgence has been almost a decade. Nevertheless, in this short period of time, deep neural networks have produced impactful and meaningful results that were never seen with more conventional machine learning methods for a long time. 
	
	Transitioning from conventional machine learning, the first attempts to use deep learning for automatic sleep staging mainly employed simple networks in traditional fashion where a short input contexts of one to a few sleep epochs around a target epoch is used to predict the sleep stage of the target epoch. Expectedly, an influx of different variants of typical standalone network architectures, such as DNN \cite{Dong2017,Wei2018}, CNN \cite{Tsinalis2016,Vilamala2017, phan2018c,Malafeev2018,Biswal2017,Sun2017,Chambon2018,Supratak2017,Andreotti2018, Phan2019b, Sors2018, Andreotti2018b}, and RNN \cite{Malafeev2018,phan2018d} (e.g., long short-term memory (LSTM) \cite{Hochreiter1997} or gated recurrent unit (GRU) \cite{Cho2014}), was seen with limited success. Although these networks are able to learn useful features to represent an input, they are unable to capture long-range dependencies between sleep epochs due to the short input context. The ability of modelling long-range dependency plays an important role in improving sleep staging performance, due to the inherently slow-transition nature of the physiological processes behind sleep stages \cite{Hartmann1968, Feinberg1979}. In order to compensate for the lack of long-term modelling ability, once such a network has been trained and each epoch is encoded into an epoch-wise feature vector, an additional RNN (e.g. LSTM) \cite{Dong2017,Supratak2017,Stephansen2018, Sun2020} was separately trained in a second stage to take into account a long sequence of epoch-wise feature vectors prior to a target epoch to classify it. These hybrid networks with two-stage training initiated by Supratak \emph{et al.} \cite{Supratak2017} boosted the performance significantly and stood out from the other exiting models at the time.
	
	In fact, the positive effect introduced by long-term modelling in the above-mentioned two-stage training scheme is not a surprise. It resembles how manual scoring is done by sleep experts who normally need to attend to a much larger context around a target epoch in order to determine its label \cite{berry_aasm_2016}. From modelling perspective, this has commonly been accomplished by Hidden Markov Models (HMM) (see \cite{Ghimatgar2020} for example) before the evolution of deep learning. However, the early works \cite{Dong2017,Supratak2017,Stephansen2018, Sun2020} came with some limitations. First, the independent two-stage training of two subnetworks is sub-optimal since it does not account for the interaction between the epoch-wise feature-learning network and the sequential-modelling counterpart, let alone its inconvenience. Second, even the sequential-modelling network (\emph{i.e.}, the bidirectional RNN) is structured to receive a sequence of epochs as input, it classifies only one target epoch at a time which is usually the last epoch in the input sequence. That is, it is tasked to encode the left-side context of the target epoch in order to make a prediction. Olesen \emph{et al.} in \cite{Olesen2021} showed that this left-side context often results in lower accuracy than when more balanced one is used.
	
\begin{figure*} [t]
		\centering
		\includegraphics[width=0.575\linewidth]{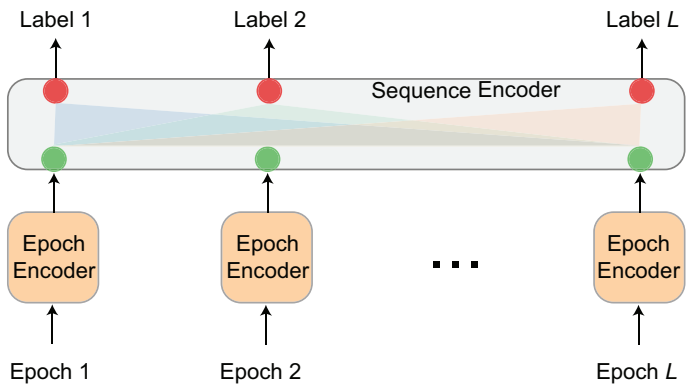}
		\caption{A schematic diagram of sequence-to-sequence sleep staging. Effectively, different epochs in the input sequence is influenced by different contexts, illustrated by the shaded regions in the sequence encoder block, depending on their absolute position in the sequence. The epoch encoder plays the role of the epoch-wise feature extractor that transforms an epoch into a feature vector representation, illustrated by a green circle. The sequence encoder enriches the presentation, illustrated by a red circle, by incorporating interaction of each epoch with other epochs in its context.}
		\label{fig:seq2seq}
	\end{figure*}

	Nevertheless, these initial results underscore the essence of long-term modelling in automatic sleep staging. Inspired by these results, since late 2018, the community has been witnessing an influx of advanced network architectures with built-in long-term modelling capacity. These networks can be generalized neatly in a common framework, namely \emph{sequence-to-sequence} sleep staging framework \cite{Phan2021}. Formally, let us denote an input sequence of $L$ epochs as $(\mathbf{S}_1, \ldots, \mathbf{S}_L)$ where $\mathbf{S}_\ell$ is the $\ell$-th epoch, $1 \le \ell \le L$. In general, the epochs can be in any form, such as raw signals or time-frequency images and they can be single- or multi-channel. A network adhering to the framework typically consists of two main components: the epoch encoder $\mathcal{F}_{E}$ and the sequence encoder $\mathcal{F}_{S}$ as illustrated in Figure \ref{fig:seq2seq}. The epoch encoder $\mathcal{F}_{E}: \mathbf{S} \mapsto \mathbf{x}$ acts as an epoch-wise feature extractor which transforms an input epoch $\mathbf{S}$ in the input sequence into a feature vector $\mathbf{x}$ for representation. As a result, the input sequence is transformed into a sequence of feature vectors $(\mathbf{x}_1, \ldots, \mathbf{x}_L)$ (represented by the green circles in the figure). Of note, $\mathcal{F}_{S}$ can be a hard-coded hand-crafted feature extractor, however, in the deep learning context, it is oftentimes a neural network (e.g., a CNN or an RNN) that learns the feature presentation $\mathbf{x}$ automatically from low-level input signals. In turn, at the sequence level, the sequence encoder $\mathcal{F}_{S}: (\mathbf{x}_1, \ldots, \mathbf{x}_L) \mapsto (\mathbf{z}_1, \ldots, \mathbf{z}_L)$ transforms the sequence $(\mathbf{x}_1, \ldots, \mathbf{x}_L)$ into another sequence $(\mathbf{z}_1, \ldots, \mathbf{z}_L)$ (represented by the red circles in the figure). In intuition, $\mathbf{z}_\ell$ is a richer representation for the $\ell$-th epoch than $\mathbf{x}_\ell$ as it not only encompasses information of the epoch but also encodes its interaction with other epochs in the sequence. More specifically, $\mathbf{z}_\ell$ is derived from $\mathbf{x}_\ell$, taking into account the left context $(\mathbf{x}_1, \ldots, \mathbf{x}_{\ell-1})$ and the right context $(\mathbf{x}_{\ell+1}, \ldots, \mathbf{x}_L)$. Eventually, the vectors $\mathbf{z}_1, \ldots, \mathbf{z}_L$ are used for classification purpose to obtain the sequence of predicted sleep stages, one for each epoch in the input sequence. 
	
	The framework features two advantages, making them overcome the limitations of the earlier proposals in \cite{Dong2017,Supratak2017,Stephansen2018, Sun2020}. First, both $\mathcal{F}_{E}$ and $\mathcal{F}_{S}$ are optimized jointly in an \emph{end-to-end} training fashion, allowing the interaction of the two network components. Second, they are tasked to solve a sequence-to-sequence classification problem, \emph{i.e.}, sequence-to-sequence sleep staging. In other words, a network classifies all the epochs in an input sequence at once rather than only targeting the last epoch. Due to this sequence-to-sequence scheme, different epochs in the input sequence is, in essence, influenced by different contexts depending on their absolute position in the sequence. This is illustrated by the shaded regions in Figure \ref{fig:seq2seq}. Leveraging this property, sampling and advancing the sequence by one epoch at a time will result in $L$ decisions for a particular epoch. These decisions are associated with diverging contexts; thus, forming an ensemble from them has been shown to lead to performance improvement \cite{Phan2019a, Phan2021a}.
	
	In Table \ref{tab:methods}, we give an overview of automatic sleep staging systems that possess capacity of long-term context modelling. The systems are presented in chronological order. On the one hand, most of the systems expectedly exploited CNN, the cornerstone of deep learning algorithms, for the epoch encoder. The spectrum of the CNN architectures varies from a very basic one \cite{Supratak2017,Seo2020} to specialized ones, such as ResNet \cite{Olesen2021}, U-Net \cite{Perslev2019,Perslev2021}, and U$^2$-Net \cite{Jia2021}. Epoch-wise features can also be learned by capturing sequential information within 30-second signals  using RNNs solely (\emph{e.g.}, LSTM \cite{Hochreiter1997} and GRU \cite{Cho2014}) \cite{Phan2019a, Guillot2020, guillot_robustsleepnet_2021} or hybrid networks (\emph{e.g.}, CRNN \cite{Seo2020, Neng2021}). Emerging network architectures like graph convolutional network (GCN) \cite{Jia2020} and Transformer \cite{Phan2021b} have also been shown to be useful for epoch encoding. On the other hand, RNNs have primarily been employed for the sequence encoder due to its well-established capability in sequential modelling. However, inter-epoch sequence modelling can also be accomplished by non-recursive architectures, such as dilated CNN \cite{Jia2021}, self-attention \cite{Eldele2021}, and Transformer \cite{Phan2021b}. It should be noted that not all of the networks in the table are strictly sequence-to-sequence (\emph{e.g.}, DeepSleepNet \cite{Supratak2017}, Stephansen \emph{et al.} \cite{Stephansen2018}, and GraphSleepNet \cite{Jia2020}) or end-to-end (\emph{e.g.}, DeepSleepNet \cite{Supratak2017}, Stephansen \emph{et al.} \cite{Stephansen2018} and Sun \emph{et al.} \cite{Sun2020}). However, in principle, they can be framed into the sequence-to-sequence framework and trained end-to-end as what was done with the end-to-end sequence-to-sequence variant of DeepSleepNet in \cite{Phan2019a}.
	
	We also collate the performance of the systems on common public sleep databases in Table \ref{tab:methods}. These results are obtained either from the original works or in other works where the systems are evaluated. We can see a Cohen's kappa of $\ge0.81$, \emph{i.e.}, ``almost perfect'' agreement level according to  the interpretation of Cohen's kappa \cite{Cohen1960}, achieved on databases with a majority of healthy subjects, for example, EDF-20, MASS, SHHS, DOD-H, DOD-O, and CHAT. However, the performance is still substandard on databases associated with sleep pathologies, such as ISRUC and CAP. It is worth stressing that the performance presented in the table should not be used out-of-context to justify a network's efficacy or compare one to another. The rational is that the potential discrepancies in evaluation setup (\emph{e.g.}, data subsets, the number of channels, etc.) and modelling (\emph{e.g.}, scratch training vs. domain adaptation, supervised vs. unsupervised learning, etc.), renders such a comparison meaningless.

	\begin{landscape}
		\setlength\tabcolsep{1.5pt}
		\begin{table*}[!t]
			\caption{Automatic sleep staging systems that has capacity of long-term context modelling published since late 2018 and sorted in chronological order. The reported performance are also presented in term of Cohen's kappa \cite{Cohen1960}. Alternatively, macro F1-score (indicated with the subscript $^{\text{f}}$) and overall accuracy (indicated with the subscript $^{\text{a}}$) are presented where Cohen's kappa is not available. Note that not all of networks here are strictly sequence-to-sequence and/or end-to-end.}
			\tiny
			\label{tab:methods}
			\begin{center}
				\begin{tabular}{|>{\arraybackslash}m{1.1in}|>{\centering\arraybackslash}m{0.25in}|>{\centering\arraybackslash}m{0.85in}|>{\centering\arraybackslash}m{0.6in}|>{\centering\arraybackslash}m{0.75in}||>{\centering\arraybackslash}m{0.35in}|>{\centering\arraybackslash}m{0.35in}|>{\centering\arraybackslash}m{0.35in}|>{\centering\arraybackslash}m{0.35in}|>{\centering\arraybackslash}m{0.35in}|>{\centering\arraybackslash}m{0.35in}|>{\centering\arraybackslash}m{0.35in}|>{\centering\arraybackslash}m{0.3in}|>{\centering\arraybackslash}m{0.3in}|>{\centering\arraybackslash}m{0.3in}|>{\centering\arraybackslash}m{0.3in}|>{\centering\arraybackslash}m{0.3in}|>{\centering\arraybackslash}m{0.3in}|>{\centering\arraybackslash}m{0.325in}|>{\centering\arraybackslash}m{0in} @{}m{0pt}@{}}
					\cline{1-19}
					Network & Year & Input & \makecell{Epoch\\Encoder} & \makecell{Sequence\\Encoder} & EDF-20 \cite{Kemp2000} & EDF-78 \cite{Kemp2000} & MASS \cite{Oreilly2014} & Physio-2018 \cite{Ghassemi2018}& SHHS \cite{Quan1997} & DOD-H \cite{Guillot2020} & DOD-O \cite{Guillot2020} & ISRUC \cite{Khalighi2016} & CAP \cite{Terzano2002}& SVUH-UCD \cite{Goldberger2000} & MESA \cite{Chen2015} & MrOS \cite{Blackwell2011,Song2015} & CHAT \cite{Marcus2013,Redline2011} & Other  \parbox{0pt}{\rule{0pt}{1ex+\baselineskip}} \\ [0ex]  	% [1ex] adds vertical spac
					
					\cline{1-19}
					DeepSleepNet \cite{Supratak2017, Phan2019a} & 2017$^*$ & Raw & CNN  & RNN & 0.760 & 0.702 & 0.800 & $-$ & $-$ & 0.843 & 0.804 & $-$ & $-$ & $-$ & $-$ & $-$ & 0.848 &$-$ \parbox{0pt}{\rule{0pt}{1ex+\baselineskip}} \\ [0ex]  	% [1ex] adds vertical
					SeqSleepNet \cite{Phan2019a} & 2018 &  Time-freq. & RNN & RNN  & 0.809 & 0.776 & 0.815 & 0.733 & 0.838 & 0.804 & 0.772 & $-$ & $-$ & $-$ & $-$ & $-$ & 0.854 & $-$ \parbox{0pt}{\rule{0pt}{1ex+\baselineskip}} \\ [0ex]  	% [1ex] adds vertical
					
					Stephansen \emph{et al.} \cite{Stephansen2018} & 2018 &  Corr. encoding & CNN & RNN  & $-$ & $-$ & $-$ & $-$ & $-$ & $-$ & $-$ & $-$ & $-$ & $-$ & $-$ & $-$ & $-$ & 0.868$^{\text{a}}$ \parbox{0pt}{\rule{0pt}{1ex+\baselineskip}} \\ [0ex]  	% [1ex] adds vertical
					
					Biswal \emph{et al.} \cite{Biswal2018a} & 2018 & Time-freq. & CNN & RNN & $-$ & $-$ & $-$ & $-$ & $-$ & $-$ & $-$ & $-$ & $-$ & $-$ & $-$ & $-$ & $-$ & 0.805 \parbox{0pt}{\rule{0pt}{1ex+\baselineskip}} \\ [0ex]  	% [1ex] adds vertical
					SleepEEGNet \cite{MousaviI2019} & 2019 & Raw & CNN & RNN & 0.790 & 0.730 & $-$ & $-$ & $-$ & $-$ & $-$ & $-$ & $-$ & $-$ & $-$ & $-$ & $-$ & $-$ \parbox{0pt}{\rule{0pt}{1ex+\baselineskip}} \\ [0ex]  	% [1ex] adds vertical		
					SimpleSleepNet \cite{Guillot2020} & 2019 & Time-freq. & RNN & RNN & $-$ & $-$ & $-$ & $-$ & $-$ & 84.6 &  82.3 & $-$ & $-$ & $-$ & $-$ & $-$ & $-$ & $-$ \parbox{0pt}{\rule{0pt}{1ex+\baselineskip}} \\ [0ex]  	% [1ex] adds vertical
					
					Chen \emph{et al.} \cite{Chen2019}$^\dagger$ & 2019 & Raw & CNN & RNN, CRF & $0.820$ & $-$ & $-$ & $-$ & $-$ & $-$ & $-$ & $-$ & $-$ & $-$ & $-$ & $-$ & $-$ & $0.670$ \parbox{0pt}{\rule{0pt}{1ex+\baselineskip}} \\ [0ex]  	% [1ex] adds vertical
					
					IITNet \cite{Seo2020} & 2019 &  Raw & CRNN & RNN & 0.780 & 0.790 & $-$ & $-$ & 0.810 & $-$ & $-$ & $-$ & $-$ & $-$ & $-$ & $-$ & $-$ & $-$ \parbox{0pt}{\rule{0pt}{1ex+\baselineskip}} \\ [0ex]  	% [1ex] adds vertical
					U-Time \cite{Perslev2019}/U-Sleep \cite{Perslev2021} & 2019 & Raw & U-Net & CNN & 0.790$^{\text{f}}$ & 0.760$^{\text{f}}$ & 0.800$^{\text{f}}$ & 0.770$^{\text{f}}$ & 0.800$^{\text{f}}$ & 0.820$^{\text{f}}$ & 0.790$^{\text{f}}$ & 0.770$^{\text{f}}$ &  0.680$^{\text{f}}$ & 0.730$^{\text{f}}$ & 0.790$^{\text{f}}$ & 0.770$^{\text{f}}$ & 0.850$^{\text{f}}$ & 0.850$^{\text{f}}$ \parbox{0pt}{\rule{0pt}{1ex+\baselineskip}} \\ [0ex]  	% [1ex] adds vertical
					TinySleepNet \cite{Supratak2020} & 2020 & Raw & CNN & RNN & 0.800 & 0.770 & 0.782 & $-$& $-$&$-$ &$-$ & $-$ & $-$ & $-$ & $-$ & $-$ & $-$ & $-$ \parbox{0pt}{\rule{0pt}{1ex+\baselineskip}} \\ [0ex]  	% [1ex] adds vertical
					GraphSleepNet \cite{Jia2020} & 2020 & Raw & GCN & Attention & $-$ & $-$ & 0.834 & $-$ & $-$ & $-$ & $-$ & $-$ & $-$ & $-$ & $-$ & $-$ & $-$ & $-$ \parbox{0pt}{\rule{0pt}{1ex+\baselineskip}} \\ [0ex]  	% [1ex] adds vertical
					Olesen \emph{et al.} \cite{Olesen2021} & 2020 & Raw & ResNet & RNN & $-$ & $-$ & $-$ & $-$ & 0.871$^{\text{a}}$ & $-$ & $-$ & 0.740$^{\text{a}}$ & $-$ & $-$ & $-$ & 0.864$^{\text{a}}$ & $-$ & 0.864$^{\text{a}}$ \parbox{0pt}{\rule{0pt}{1ex+\baselineskip}} \\ [0ex]  	% [1ex] adds vertical
					Jaoude \emph{et al.} \cite{Jaoude2020} & 2020 & Raw & CNN & RNN & $-$ & $-$ & $-$ & $-$ & $-$ & $-$ & $-$ & $-$ & $-$ & $-$ & $-$ & $-$ & $-$ & 0.740 \parbox{0pt}{\rule{0pt}{1ex+\baselineskip}} \\ [0ex]  	% [1ex] adds vertical
					Sun \emph{et al.} \cite{Sun2020} & 2020 & Raw, hand-crafted & CNN & RNN & $-$ & $-$ & 0.795 & $-$ & $-$ & $-$ & $-$ & $-$ & $-$ & $-$ & $-$ & $-$ & $-$ & $-$ \parbox{0pt}{\rule{0pt}{1ex+\baselineskip}} \\ [0ex]  	% [1ex] adds vertical
					
					Korkalainen \emph{et al.} \cite{Korkalainen2020} & 2020 & Raw & CNN & RNN & $-$ & $0.780$ & $-$ & $-$ & $-$ & $-$ & $-$ & $-$ & $-$ & $-$ & $-$ & $-$ & $-$ & $0.790$ \parbox{0pt}{\rule{0pt}{1ex+\baselineskip}} \\ [0ex]  	% [1ex] adds vertical
					
					Qu \emph{et al.} \cite{Qu2020} & 2020 & Raw & CNN, ResNet & Self-attention & $0.780$ & $-$ & $0.800$ & $-$ & $-$ & $-$ & $-$ & $-$ & $-$ & $-$ & $-$ & $-$ & $-$ & $-$ \parbox{0pt}{\rule{0pt}{1ex+\baselineskip}} \\ [0ex]  	% [1ex] adds vertical
					
					HNSleepNet \cite{Chen2020} & 2020 & Raw & CNN & RNN, Attention & $0.780$ & $-$ & $0.810$ & $-$ & $-$ & $-$ & $-$ & $-$ & $-$ & $-$ & $-$ & $-$ & $-$ & $-$ \parbox{0pt}{\rule{0pt}{1ex+\baselineskip}} \\ [0ex]  	% [1ex] adds vertical
					
					Li \emph{et al.} \cite{Li2020} & 2020 & Raw & CNN & RNN, Attention & $0.790$ & $-$ & $-$ & $-$ & $-$ & $-$ & $-$ & $-$ & $-$ & $-$ & $-$ & $-$ & $-$ & $-$ \parbox{0pt}{\rule{0pt}{1ex+\baselineskip}} \\ [0ex]  	% [1ex] adds vertical
					
					FCNN+RNN \cite{Phan2021a} & 2020 & Raw & Fully CNN & RNN & 0.775 & 0.759 & 0.806 & 0.738 & 80.9 & $-$ & $-$ & $-$ & $-$ & $-$ & $-$ & $-$ & 0.847 & $-$ \parbox{0pt}{\rule{0pt}{1ex+\baselineskip}} \\ [0ex]  	% [1ex] adds vertical
					XSleepNet \cite{Phan2021a} & 2020 & Raw, time-freq. & CNN, RNN & RNN & 0.813 & 0.778 & 0.823 & 0.746 & 0.847 & $-$ & $-$ & $-$ & $-$ & $-$ & $-$ & $-$ & 0.857 & $-$ \parbox{0pt}{\rule{0pt}{1ex+\baselineskip}} \\ [0ex]  	% [1ex] adds vertical
					RobustSleepNet \cite{guillot_robustsleepnet_2021} & 2021 & Time-freq. & RNN & RNN & 0.817$^{\text{f}}$ & 0.779$^{\text{f}}$ & 0.825$^{\text{f}}$ & $-$ & 0.800$^{\text{f}}$ & 0.851$^{\text{f}}$ & 0.827$^{\text{f}}$ & $-$ & 0.738$^{\text{f}}$ & $-$ & 0.795$^{\text{f}}$ & 0.756$^{\text{f}}$ & $-$ & $-$ \parbox{0pt}{\rule{0pt}{1ex+\baselineskip}} \\ [0ex]  	% [1ex] adds vertical
					CCRRSleepNet \cite{Neng2021} & 2021 & Raw  & CRNN & RNN & 0.780 & $-$ & $-$ & $-$ & $-$ & $-$ & $-$ & $-$ & $-$ & $-$ & $-$ & $-$ & $-$ & $-$ \parbox{0pt}{\rule{0pt}{1ex+\baselineskip}} \\ [0ex]  	% [1ex] adds vertical
					Eldele \emph{et al.} \cite{Eldele2021} & 2021 & Raw  & CNN & Self-attention & 0.790 & 0.740 & $-$ & $-$ & 0.780 & $-$ & $-$ & $-$ & $-$ & $-$ & $-$ & $-$ & $-$ & $-$ \parbox{0pt}{\rule{0pt}{1ex+\baselineskip}} \\ [0ex]  	% [1ex] adds vertical
					
					RecSleepNet \cite{Nie2021} & 2021 & Raw & CNN & RNN & $0.813$$^{\text{f}}$ & $0.779$$^{\text{f}}$ & $-$ & $-$ & $-$ & $-$ & $-$ & $0.779$$^{\text{f}}$ & $-$ & $0.743$$^{\text{f}}$ & $-$ & $-$ & $-$ & $-$ \parbox{0pt}{\rule{0pt}{1ex+\baselineskip}} \\ [0ex]  	% [1ex] adds vertical

					Coon \emph{et al.} \cite{Coon2021} & 2021 & Raw & CNN & RNN & $-$ & $-$ & $-$ & $-$ & $-$ & $-$ & $-$ & $-$ & $-$ & $-$ & $-$ & $-$ & $-$ & $0.769$$^{\text{a}}$ \parbox{0pt}{\rule{0pt}{1ex+\baselineskip}} \\ [0ex]  	% [1ex] adds vertical
					
					SalientSleepNet \cite{Jia2021} & 2021 & Raw & U$^2$-Net & Dilated CNN &  0.830$^{\text{f}}$ & 0.795$^{\text{f}}$ & $-$ & $-$ & $-$ & $-$ & $-$ & $-$ & $-$ & $-$ & $-$ & $-$ & $-$ & $-$ \parbox{0pt}{\rule{0pt}{1ex+\baselineskip}} \\ [0ex]  	% [1ex] adds vertical
					SleepTransformer \cite{Phan2021b} & 2021 & Time-freq. & Transformer & Transformer & $-$ & 0.789 & $-$ &  $-$ & 0.828 & $-$ & $-$ & $-$ & $-$ & $-$ & $-$ & $-$ & 0.842 & $-$ \parbox{0pt}{\rule{0pt}{1ex+\baselineskip}} \\ [0ex]  	% [1ex] adds vertical
					\cline{1-19}
				\end{tabular}
			\end{center}
			\label{tab:databases}
		\end{table*}
		\footnote[0]{\scriptsize $^*$DeepSleepNet was introduced by Supratak \emph{et al.} \cite{Supratak2017} in 2017 and the end-to-end version was presented as a baseline in the SeqSleepNet work \cite{Phan2019a} by Phan \emph{et al.} in 2018.}
		\footnote[0]{\scriptsize $^\dagger$Chen \emph{et al.} \cite{Chen2019} conducted 4-stage classification rather 5-stage classification in other works in the table.}
	\end{landscape}

	\section{Challenges and future directions}
	
	In our view, automatic sleep staging with PSG on healthy people has basically been solved, not only for adults \cite{Biswal2018a, Phan2021a} but also for children \cite{Phan2021c}. The comparative study by Phan \emph{et al.} \cite{Phan2021c} showed that different network architectures under the same sequence-to-sequence framework result in a similar ``almost perfect'' consensus level (according to the interpretation of Cohen's kappa \cite{Cohen1960}) and little discrepancy was seen among their staging outcomes. This suggests that there is probably little room for accuracy improvement within the same sequence-to-sequence framework. Furthermore, the improvement, if any, is not necessarily meaningful.
	
\begin{figure*} [!t]
		\centering
		\includegraphics[width=0.85\linewidth]{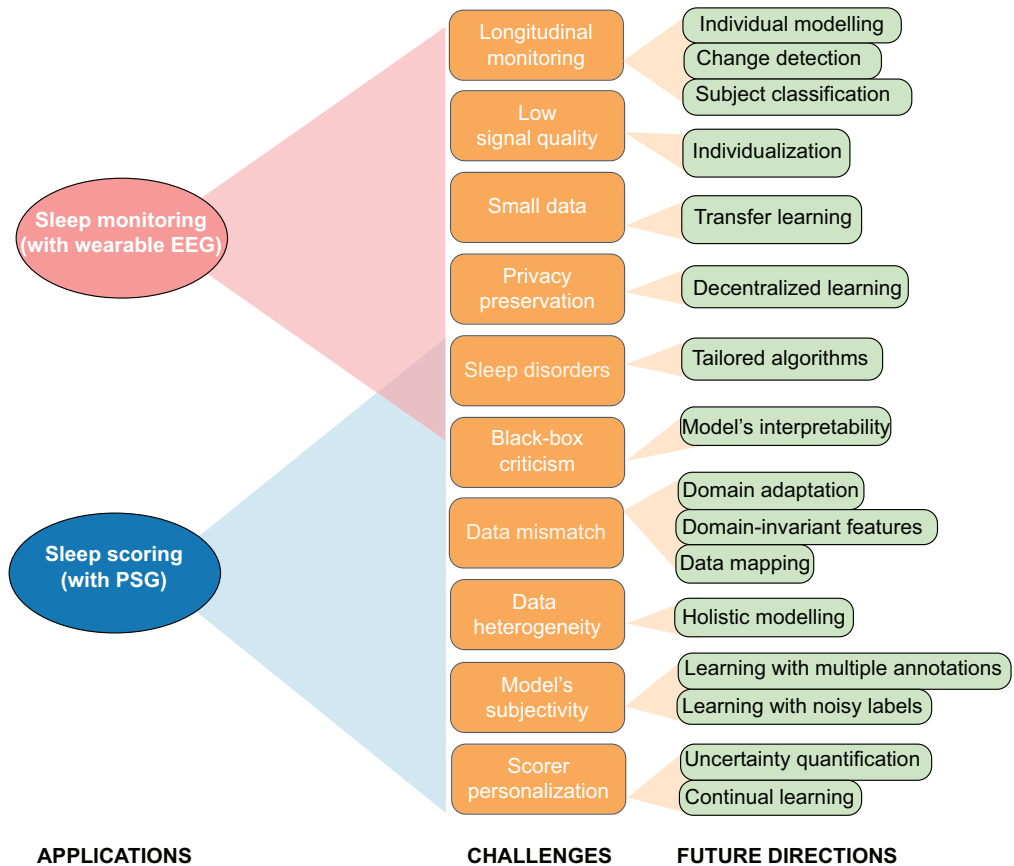}
		\caption{Applications, challenges, and directions in automatic sleep staging.}
		\label{fig:challenges}
	\end{figure*}

		While automatic sleep scoring of PSG recordings has come a very long way as described above, there are still challenges to be overcome. These, of course, should be viewed as opportunities for innovation. In Figure \ref{fig:challenges}, we give an overview of the challenges around two critical applications, sleep scoring with PSG in clinical spaces and sleep monitoring with wearable EEG in daily living environments, and directions for future works to address the challenges. Before discussing the individual challenges, we feel it is valuable to highlight the overarching contexts of the two applications:
	
		\emph{Clinical PSG scoring:} It is not sufficient to have high quality PSG scoring of healthy people. In a clinical setting, the tools applied should be equally capable when confronted with non-textbook sleep phenotypes, where the sleep EEG may either be masked by disease-related artifacts, or where the sleep EEG itself may be drastically changed by the patient’s condition that a correct sleep scoring either requires specialized routines or may even be impossible. An automatic sleep scoring algorithm must be able to handle this situation transparently and reliably. Thus, to obtain more widespread adoption clinically, automatic sleep scoring should be as robust as manual scoring, and deliver outputs which are easy to fit into clinical workflows. In Figure \ref{fig:challenges}, this relates particularly to `data heterogeneity', `model interpretability', `learning with noisy labels' and `tailored algorithms'. 
	
		\emph{Wearable EEG scoring:} Medical grade mobile sleep monitoring has great potentials for revolutionizing healthcare both in screening, diagnosing and follow-up. A high quality monitoring platform would allow easy recording of weeks of sleep from each individual, without incurring higher healthcare costs or discomfort the patient. Such data would be much more representative of the patient’s actual sleeping patterns, and alleviate the present issues with phenomena which are only periodic, or which may be impacted by patients sleeping in unfamiliar environments. To reap the full benefits from such a monitoring device, we need analysis tools which can be tailored to the individual, which can detect subtle changes in sleep patterns, and which can describe a patient’s sleep in different, quantitative terms than the present single-night hypnograms. 
	
	The discussion below is structured following the overview outlined in Figure \ref{fig:challenges}.
	
	\subsection{Longitudinal monitoring}
	
	A particularly interesting development in sleep monitoring, closely related to automatic scoring, is the feasibility of long term monitoring. With conventional PSG setups, scored by hand, it is generally too expensive, not to say inconvenient for the wearer, to perform nightly recordings for weeks on end. However, when scoring is free, and when the hardware can be self-applied \cite{mikkelsen_accurate_2019, arnal_dreem_2019, miettinen_success_2018} and is relatively unobtrusive, longitudinal monitoring becomes much more feasible. Interested readers are recommended to refer to \cite{Imtiaz2021} for a complete review on devices for wearable sleep staging. These types of devices, and the data sets recorded with them, open new, very interesting avenues of research:
	
	\subsubsection{Individual modelling}
	\label{sssec:individual_modelling}
	Given a large number of nights from an individual, it seems both possible and advantageous to create personal models using this data. Indeed, we may be inspired by more general approaches for semi-supervised machine learning \cite{kang_contrastive_2019,sun_unsupervised_2019,padmanabhan_speaker_1998}. We could imagine that a successful approach would not only lead to an individual model, but a model which could keep up with the possible ``concept drift'' that is caused by the long-term changes in a person's sleep patterns. At the same time, such an adaptive approach should be resistant to ``catastrophic forgetting'' \cite{french_catastrophic_1999}, in which the sequential learning of a neural network causes it to forget how to solve previous problems, in this case how to deal with types of sleep or events which only happen very infrequently for the given subject. We suspect that simply requiring a long ``memory''  is sufficient to solve this problem. Given the performance of models trained as leave-one-subject-out, the size of standard data sets, the fact that inter-subject variation is estimated to be a significant driver of sleep data variation \cite{buckelmuller_trait-like_2006,finelli_individual_2001,hemmsen_long-term_2021,tucker_trait_2007} and studies evaluating the performance of individualized models \cite{mikkelsen_accurate_2019,phan_personalized_2020}, we expect that on the order of a 100 nights would be sufficiently long.
	
	\subsubsection{Change detection}
	
	A natural extension to this discussion is detection of sudden (between nights), serious changes to an individual's sleep. Given the large variation in sleep in a given subject, this is not readily feasible based on a few nights. However, based on perhaps weeks of data before and after a significant event (\emph{e.g.}, surgery, change in medication, disease onset, etc.), it seems possible that we could reliably detect that a patient had started to sleep differently. Conceivably, such change detection will be aided by the development of a continuously updating model (since this should entail detecting the need for significant updates). 
	
	\subsubsection{Subject classification}	
	
	It has been shown that despite the significant night-to-night variation in an individual's sleep, it is possible to define ``trait-like characteristics'', special to the individual \cite{finelli_individual_2001,tucker_trait_2007,chua_individual_2014,hemmsen_long-term_2021}. Ideally, this could mean that given a sufficient amount of sleep recordings from an individual, those could be transformed into a reliable biomarker, which could be used to determine not only changes in sleep patterns, but also whether a given patient's sleep was indicative of certain specific diseases, such as RBD. 
	
	\subsection{Low signal quality - device limitations for mobile sleep monitoring}
	
	Mobile sleep monitoring devices have great potential for helping both healthy and sick users get increased knowledge about their own sleep. However, as has been shown in multiple studies \cite{mikkelsen_accurate_2019,arnal_dreem_2019}, even the best studies do not achieve the same inter-scorer reliability measures as PSG-based approaches (state-of-the-art values for Cohen's kappa seem to be about 0.75 for mobile devices \cite{arnal_dreem_2019,Mikkelsen2021}, while PSG data leads to values above 0.8).  A lower signal-to-noise ratio for these types of data (relative to PSG recordings) is likely the main cause for the degraded performance in mobile solutions. Multiple studies \cite{mikkelsen_machine-learning-derived_2018,Mikkelsen2021} have achieved much better performance from the same number of PSG nights as mobile EEG nights (using concurrent recordings). This means that while small data sets can definitely be an issue as discussed in Section \ref{ssec:small_database}, it seems that making many recordings is likely not sufficient. We can model this phenomenon by imagining the mobile data set as a projection of the high-dimensional PSG data into a lower-dimensional space, with resulting information loss. Given the differences in neuroanatomy, it is reasonable to consider this projection to be subject dependent, and depending on the device design, we could also expect there to be a variation between recording days (because the device may be mounted differently each time). 
	
	Several studies have shown that including subject specific information increases sleep scoring performance \cite{mikkelsen_accurate_2019,mikkelsen_automatic_2017,phan_personalized_2020}, which is also in line with general EEG studies showing significant differences between individuals \cite{Palaniappan2007,mikkelsen_eegs_2021}. Mikkelsen \emph{et al.} \cite{mikkelsen_accurate_2019} found that the differences between algorithms trained on data just from the same subject, and data from both the same subject and many other subjects were minimal. This indicates that the benefit from personalizing sleep scoring algorithms is not that irrelevant data is excluded, but rather that maximally relevant data is included. If this trend were to scale to much larger cohorts, one could imagine that for sufficiently large cohorts, the personalized and general algorithms would perform similarly. However, achieving such a broad training set may be unrealistic in practice. 
	
	Individualization of algorithms have been done in multiple ways. Some studies have used random forest ensemble models \cite{mikkelsen_automatic_2017,mikkelsen_accurate_2019,gangstad_automatic_2019}, which can be trained using very little data. This makes it possible to create full sleep scoring models using only a single or few nights of data from an individual. Other studies, using deep neural networks, have instead resorted to variations of fine tuning population models to individuals. Phan \emph{et al.} \cite{phan_personalized_2020} explored a technique where the model, during fine tuning, was penalized for making large changes to the output in the source domain, effectively limiting the risk of overfitting to the fine tuning data set.
	
	Note that while some groups focus on developing the entire device, others have strictly worked on developing sleep scoring algorithms for generic ``single channel EEG'' data sets, without relating it to a specific device \cite{Koley2012,olesen_deep_2020}. In this discussion we have lumped the two approaches together. We note also that similar observations, regarding mobile device monitoring, are made in the review by Chriskos \emph{et al.} \cite{chriskos_review_2021}. 
	
	\subsection{Modelling with a small amount of data}
	\label{ssec:small_database}
	Training a deep neural network generally requires a large amount of data. In fact, deep-learning based sleep staging models only reach expert-level performance when the training cohort is large, \emph{i.e.}, hundreds or thousands of subjects \cite{Phan2019a, Phan2021a, Biswal2018a, Perslev2021}. The networks trained  with a small cohort continue exhibiting substandard performance. Unfortunately, in practice, many sleep studies only have access to a small cohort, in the order of a few dozens of subjects, for example when studying a particular sleep disorder \cite{Cooray2019, Andreotti2018}. 
	
	This scenario is particularly common in studies exploring the feasibility of a new monitoring device, for example mobile EEG devices \cite{mikkelsen_accurate_2019, mikkelsen_machine-learning-derived_2018, heremans2021}. While the PSG benefits from being an established standard with an enormous user base, new alternatives, by definition, do not. Add to this that many such devices will undergo multiple generations which may not be compatible. Finally and crucially, new training sets will usually require special recordings of both a device and PSG signals, to obtain the necessary ground truth manual PSG scoring which constitutes the training labels. All together, this means that algorithms for new sleep monitoring devices usually have to be trained with quite small data sets.  
	
	The most popular solution to this problem is to use transfer learning, usually in the sense that a neural network is trained on a large sleep data set, often consisting of PSG recordings. This model is then fine-tuned using the new data set \cite{Phan2019f,Phan2021,guillot_robustsleepnet_2021,olesen_deep_2020, heremans2021}. Although model fine tuning often results in better performance than scratch model training, the gains are modest in some cases. The problem is that by fine tuning, we essentially further train a pretrained model with a small amount of data that easily causes overfitting without a proper regularization, especially when the source domain and the target domain are significantly different. This can be remedied by fine tuning just a part of the  pretrained model, however, identifying the most relevant layers for fine tuning still remains an unsolved question. EEG data augmentation \cite{Fan2020} is another direction to explore.

	\subsection{Privacy preservation: a note for sleep monitoring}	
	%privacy: 
	%- it should be able to run locally, so not huge models
	Brain waves are a rich source of information from which deciphering numerous sensitive piece of information has been shown possible, such as identity \cite{Palaniappan2007}, age \cite{Zoubi2018}, gender \cite{Wang2019}, emotion \cite{Alarcao2019}, preference \cite{Sangnark2021}, personality \cite{Zhao2018}, etc. This poses a challenge to protect this information from user perspective and to comply with legal restrictions, such as General Data Protection Regulation (GDPR) \cite{gdpr} in European Union and Consumer Privacy Bill of Rights (CPBR) \cite{Gaff2014} in the US.
	
	Traditionally, when sleep staging models are trained, adapted, and deployed centrally, EEG signals are expected to be sent via some communication means. Data privacy and security concerns should be paid heed in this case, however, it is beyond the scope of this article. From the algorithmic perspective, federated learning \cite{McMahan2017} emerges as a promising solution to address the fundamental problems of privacy and locality of data in the conventional centralized setting. Instead of bringing data to the (centralized) code, federated learning brings the code to (decentralized) data and exploits the distributed resources to train a model collaboratively. Thus, it gets rid of the need for the data to be transferred to a single server. Several open-source federated learning systems are available, e.g.  FATE \cite{Webank}, PaddleFL \cite{Baidu}, TensorflowFL \cite{Google}, and Pysyft \cite{OpenMined}, that would facilitate future research in this direction. Developing compact deep neural networks \cite{Xia2020} for sleep staging also becomes relevant. This particularly fits to longitudinal monitoring as model personalization and development can be done on devices, like wearables, smartphones, or Internet-of-Things (IoT) edge devices and the person's data can stay local. Since these devices operate under run-time energy and memory storage constraints, they can only accommodate compact models due to their reduced energy consumption, memory requirement, and inference latency. First, compact versions of existing state-of-the-art models can be derived via quantization \cite{Hubara2018, Zhu2017} to reduce bit-depth of the weights and pruning \cite{Han2016, Molchanov2017} to remove redundant weights. Second, a network architecture can be hand-designed to remove redundancy, for example depth-wise convolution \cite{Ma2018} and shift-based module \cite{Wu2018}, and thus improve efficiency. Although this approach often results in good model compactness, there is no guiding principle in hand-engineering a network architecture and most of the existing works are more or less based on trial-and-error. An alternative approach is to automatically search for network architectures, \emph{i.e.}, neural architecture search (NAS) \cite{Baker2017,Zoph2017}, which has seen some success, for example in the image domain \cite{Tan2019}

	\subsection{Disorders affecting sleep structure}
	
	If a subject suffers from severe neurological disorders, this can have a drastic impact on their sleep. It may both change the structure, timing and outward characteristics of the individual's sleep \cite{iranzo_sleep_2016}, and it can also change the physiological features of the various sleep stages \cite{santamaria_scoring_2011}. This can eventually lead to different types of issues, which require different types of solutions. First, if the manual scoring of the recordings becomes harder, training and validation of a sleep-staging model becomes equally hard \cite{Stephansen2018,Korkalainen2020}. We are not aware of any methodological studies on how to deal with this issue in sleep scoring, but it is a general issue for most types of medical data. Certainly, some of the methods employed in, for instance, medical imaging analysis \cite{tajbakhsh_embracing_2020} could be re-purposed for sequence-to-sequence sleep-staging models. Second, if the underlying issue is that distinct sleep stages are becoming less defined (which could be the case in very advanced brain damage), it is possible that more flexible approaches such as continuous sleep depth estimation \cite{asyali_determining_2007,carrubba_continuous_2012} are a better fit. Third, even if the above issues do not appear, the changes in transition probabilities or even feature spaces have to be absorbed by a classification algorithm to achieve good sleep-staging performance. Fortunately, some studies have shown that a model's performance can be improved significantly if it is individualized, for instance when the subjects are suffering from epilepsy \cite{gangstad_automatic_2019} or RBD \cite{Andreotti2018}.
	
	A particular implementation which is interesting in this context is the ``ASEEGA'' algorithm \cite{berthomier_automatic_2007} which extracts frequency-based features and scales them according to the individual night. This results in an algorithm which achieves an average Cohen's kappa of 0.8 for healthy adults and full PSG setups (thus, not quite state-of-the-art), but which, on the other hand, appears to be quite resistant to perturbations caused by sleep disorders \cite{peter-derex_automatic_2021} (reaching Cohen's kappa values between 0.75 and 0.80, depending on disorders). %It is a shame that the code appears to be closed-source. 
	Another promising approach, similar to the ``ASEEGA'' algorithm, is to feed an entire night into the algorithm at once. In theory, this could enable the algorithm to detect subject- or disease-specific perturbations, and adjust for them. We have seen the approach applied by Li \emph{et al.} \cite{li_deepsleep_2021} for arousal detection, to great success.

	\subsection{Black-box criticism and interpretability}
	\subsubsection{Black-box criticism and adoption}
	Trust, a psychological mechanism to deal with uncertainty, is a crucial factor influencing interactions and relationships between human and AI, particularly between clinicians and AI in the healthcare domain. This is the chief mechanism that shapes the use and adoption of AI in healthcare settings where life is involved \cite{Asan2020}. With its capabilities, deep learning has demonstrated its benefits to healthcare in many aspects: superior performance, capability of handling large and complex data, data-driven learning ability, etc. Examples are the application of deep learning to image-based diagnosis \cite{Ting2018}, clinical outcome prediction \cite{Rajkomar2018}, automatic electrocardiogram (ECG) analysis \cite{Ribeiro2020}, automatic sleep analysis \cite{Stephansen2018}, mental health screening \cite{Su2020}, intelligent assistive technologies for dementia care \cite{Ienca2017}, to mention a few. However, the complex nature of these algorithms and their inherent ``black-box''-ness have been deterring medical professionals' trust \cite{Rajkomar2018,Lipton2018}. The way that a deep neural network processes input data through interconnected layers to arrive at its staging decisions poses difficulties in deciphering how it learns to produce the outputs. Expectedly, automatic sleep staging systems are not an exception as black-box skepticism remains one of the main questions around their clinical value and adoption.

	\subsubsection{Interpretability}
	\label{ssec:interpretability}
	Interpretability is critical for a trustworthy sleep-staging system due to the fact that sleep stages are often ambiguous and even different human experts tend to disagree to a certain extent \cite{guillot_robustsleepnet_2021,DankerHopfe2009}. While addressing this interpretability problem is mandatory to unleash the clinical value of deep-learning-based sleep staging algorithms, it requires novel technical approaches to understanding the behaviour of these AI systems (\emph{i.e.}, explainable AI \cite{Gunning2019})
	
	A few attempts have been made to introduce interpretability to a deep-learning-based automatic sleep staging system. Most of them explain the models using feature visualization methods, such as sensitivity maps \cite{Yosinski2015, Rasmussen2011} by Vilamala \emph{et al.} \cite{Vilamala2017}, Guided Gradient-weighted Class Activation Maps (Guided Grad-CAM) \cite{Selvaraju2017} by Andreotti \emph{et al.} \cite{Andreotti2018b}, saliency map \cite{Qin2020} by Jia \emph{et al.} \cite{Jia2021}, and self-attention score \cite{Vaswani2017} by Phan \emph{et al.} \cite{Phan2021b}. In another work, Lee \emph{et al.} \cite{Lee2020} proposed to associate a model's learning process with expert-defined EEG patterns. These patterns were used as templates for the first convolutional kernels of a CNN and were located in a test EEG signal via cosine similarity maximization to achieve interpretability. Al-Hussaini \emph{et al.} \cite{Al-Hussaini2019} gained interpretability from the perspective of decision rules by coupling a deep learning model with a regression tree. Prototypes in the high-dimensional embedding space of a CNN were firstly derived and used to generate similarity for each PSG epoch with the expert-defined rules. The similarity scores were then classified by a decision tree. Several resulting splitting rules of the decision tree were found similar to the guidelines for human annotators \cite{berry_aasm_2016}.
	
	Ultimately, future research towards interpretable automatic sleep staging could benefit from explainable deep learning research in general. On the one hand, new backpropagation-based methods (e.g., layer-wise relevance propagation \cite{Bach2015,Montavon2018}, Deep Learning Important FeaTures (DeepLIFT) \cite{Shrikumar2017}, and integrated gradients \cite{Sundararajan2017}) and perturbation-based methods (e.g., occlusion sensitivity \cite{Zeiler2014}, representation erasure \cite{Li2016}, meaningful perturbation \cite{Fong2017}, and prediction difference analysis \cite{Zintgraf2017}) can be explored to improve models' explanation via scientific visualization of characteristics of an input that influence the output of a model. Designing intrinsically explainable deep networks, like \cite{Jia2021, Phan2021b}, that can jointly optimize model performance and provide explanations as part of the model output is another potential direction. These intrinsic methods are probably more desirable than the post-hoc methods that seek explanation of models that were never designed to be explainable in the first place. On the other hand, there are model distillation approaches \cite{Ribeiro2016,Hinton2015} in which the knowledge encoded within a deep learning model (\emph{i.e.}, the ``black-box'' model) is distilled into a ``white-box'' model  which is meant to identify the decision rules influencing the outputs of the deep learning model, as shown in \cite{Al-Hussaini2019}. The distilled models, potentially simple and interpretable, such as decision tree or logistic regression, offer the explanation power while still achieving reasonable performance. In this way, one could alleviate the compromise between interpretability and prediction accuracy to some extent. 
	
	However, it remains an open question how the explainability of a model could be objectively quantified, evaluated, and compared. Our opinion is that an explainable AI system for automatic sleep scoring should be inspired by the way a sleep expert performs manual scoring to provide interpretability to (1) whether the features and the rules resulted from an algorithm are clinically relevant to and underpin sleep and (2) how the decision on a target epoch is made under the influence of its neighboring epochs given their strong dependency due to the continuous nature of sleep. However, answering these questions in automatic sleep staging is tricky. First, most of the existing approaches explain the models via the prism of expert-defined rules and features, however, many of these features are not well-defined while the majority of the rules for human annotators are vague \cite{berry_aasm_2016}. Second, these features and rules cannot be used in many scenarios, for instance, wearable EEG \cite{mikkelsen_accurate_2019, mikkelsen_automatic_2017, Sterr2018} since the underlying signals are different from scalp EEGs and not readily interpretable for a human scorer. Third, in practice, many features and rules learned by these networks do not conform to the established features and rules. However, these challenges point in the direction of moving beyond interpretability. We envision that efforts should also be spent on understanding the disharmonizing features and rules resulted purely from data. Clinical explanations for them would potentially help us to gain further insights about underlying neurophysiology of human sleep, which, in turn, could be used to update the manual-scoring features and rules \cite{Penzel2013, Al-Hussaini2019}.

	\subsection{Data mismatch due to distributional shifts between datasets/cohorts}
	\label{ssec:data_mismatch}
	Sleep data typically come from different sources with a wide range of institutions, demographics, diseases, modalities, devices, and acquisition conditions. As a result, these mismatches violate the data assumption of being independent and identically distributed (i.i.d.) required for a machine learning system. Even when a deep-learning sleep staging model is trained on large amounts of data, resulting in powerful hierarchical representations, the discrepancies are still computationally significant, degrading the accuracy of sleep staging models on unseen data with a shift (\emph{i.e.}, mismatch) in their distribution \cite{Phan2021}. A naive solution for this problem is to form training data from as many conditions as possible \cite{Olesen2021}, ideally from all types of conditions that will be foreseeably encountered in the deployment phase. However, this is expensive, time-consuming, and infeasible. In addition, novel setups will likely emerge in the study of particular sleep disorders \cite{Cooray2019, Stephansen2018} or when exploring the feasibility of  new monitoring devices \cite{mikkelsen_machine-learning-derived_2018, mikkelsen_automatic_2017, arnal_dreem_2019, Myllymaa2016, heremans2021}.
	
	%https://arxiv.org/pdf/1812.05313.pdf
	As the data mismatches cannot be simply reversed by signal preprocessing, approaches to migrate a pretrained model to a target cohort with an unseen condition (\emph{e.g.}, via \emph{transductive} transfer learning or domain adaptation \cite{Phan2021,Phan2019f,Eldele2021b,Yoo2021,Zhao2021,heremans2021}) have been adopted. While most existing works on this direction utilized a large labelled database for model pretraining in a supervised fashion, semi-supervised \cite{Engelen2020} and unsupervised (i.e. self-supervised) \cite{Jing2021, Jiang2021, Yang2021, Mohsenvand2020, Banville2021} training regimes would further allow  leveraging unprecedentedly large amounts of unlabelled data for this purpose. However, before migrating a pretrained model to a target domain, distributional shifts need to be detected and quantified to indicate the variation of the model  and whether the migration is necessary or not. Entropy of the probability outputs could potentially serve this purpose \cite{Mikkelsen2020}. Then, methods for model migration from a source domain to a target domain can be categorized depending on the availability of labelled data in the target domain. 
	%Furthermore, these methods are mainly \emph{transductive} transfer learning (or domain adaptation) \cite{Daume2006,Pan2010} as the classification task is typically the same in both the source and target domains, but the domain differs. 
	In the best case when all data is labelled, supervised domain adaptation methods \cite{Tan2018}, such as \cite{Phan2021, Phan2019f,heremans2021}, appear to be most sensible. In these methods, a pretrained model needs to undergo a fine tuning process, \emph{i.e.}, the model is further trained in a supervised fashion using the target domain's labelled data. When only a part of the target domain data are labelled, supervised domain adaptation is, in essence, still feasible if the amount of labelled data is sufficient. Otherwise, fine tuning using a small amount of data will be exposed to a great risk of overfitting. In any case, a better solution is to exploit semi-supervised domain adaptation methods, that incorporate semi-supervised learning (SSL) \cite{Engelen2020} and domain adaptation \cite{Tan2018}, to leverage both labelled and unlabelled data at the same time. For example, a pretrained model can be fine tuned to simultaneously minimize the sum of supervised classification and unsupervised reconstruction cost functions \cite{Rasmus2018}. In the worst case when all the data are unlabelled, unsupervised domain adaptation will be a natural choice \cite{Wilso2020}. For example, encouraging results have been reported using adversarial domain adaption methods to match the feature distributions of the source and target domains via gradient reversal from a domain classifier that is tasked to discriminate between the two domains \cite{Nasiri2020, Yoo2021, Zhao2021}. A pretrained network can be also be adapted to a target domain by modulating the domain-specific statistics of deep features stored in the network's normalization layers like batch normalization \cite{Fan2022}. While the reliance on target data labels are costly as human scoring is required, in general, the performance gains are proportional to the amount of labelled data. With the same target-domain data, the gains from supervised transfer learning methods are expected to be higher than semi-supervised ones which are in turn higher than unsupervised ones. However, success of model migration is also subject to training strategies, network-architecture choices and datasets \cite{Phan2021}.
	
	An alternative approach to the domain adaptation is data mapping. This has remained mostly unexplored for sleep data. The difference is that domain adaptation requires modification of the model parameters while data mapping aims to modify the data to map them from one domain to another. To this end, a mapping function can be learned to map from a certain target domain to the source domain. Evaluating a sleep staging model on a target domain, the test data is firstly fed to the domain mapping function to make them look like the source data before sleep staging takes place. Inspired by the success of generative adversarial networks (GAN) \cite{Goodfellow2016b} in image-to-image translation, subject-to-subject or sequence-to-sequence PSG mapping could potentially be done similarly with these GAN variants. However, the challenge here is that we may wish to achieve the mapping by modifying the traits of the mismatched factors in data while preserving the sequential nature of the sleep data.
	
	Another approach to align source and target domains is to force a sleep staging model to learn domain-invariant feature representations \cite{Zhao2019,heremans2021}. In other words, the features learned by the model follow the same distribution no matter whether the input are from the source or target domain, and so representing the underlying sleep stages while being agnostic to other factors. As a consequence, the model trained on the source domain can generalize well to the target domain without the necessity of modification of the model parameters or data mapping. One way to achieve this is to minimize the distances (\emph{e.g.}, Wasserstein distance \cite{Chambon2018b, Long2016}) between the distributions during training in addition to the classification task. An alternative to distance minimization is to incorporate reconstruction losses \cite{Ghifary2016,Bousmalis2016} to encourage the learned features to reconstruct well either the target domain data or both the source and target domain data. Another possibility is to rely on an adversarial domain classifier which is tasked to discriminate the source and target domain. In light of adversarial training as in a GAN \cite{Goodfellow2016b}, the idea is then to train the sleep staging model to learn the features such that the domain classifier is unable to distinguish from which domain the features originated \cite{Zhao2019}. However, this approach requires data from both domains to be available at the training phase.
	
	\subsection{Heterogeneity: a challenge beyond data mismatch}
	
	Apart from the data-mismatch challenge discussed in Section \ref{ssec:data_mismatch}, heterogeneity is another challenge emerging from the data originated from different sources, or even from different subjects of the same cohort. Typically the number of channels and modalities in PSG recordings varies significantly due to the differences in channel layout and recording setup. This is not a major challenge in many existing network architectures relying on a single channel of one or a few modalities (\emph{i.e.}, EEG, EOG, and EMG) \cite{Phan2019a, Supratak2017, Perslev2021, Phan2021a}. However, it limits the applicability of those utilizing a larger number of channels \cite{Jia2020, Chambon2018} when one or more employed channels are missing in test data. In practice, different channels and modalities manifest different perspectives of the underlying neurophysiological processes in human sleep. For example, Alpha rhythm appears most clearly in occipital lobe, sawtooth waves characterizing REM are best captured in central lobe and K-complex events, the hallmark of N2, are best observed in central lobe \cite{berry_aasm_2016}. As a result, consolidating information from all available channels of PSG data in a holistic view would potentially improve sleep-staging performance. This will also facilitate model building from an unprecedented amount of data gathered from different sources \cite{Perslev2021,Stephansen2018,guillot_robustsleepnet_2021}.
	
	Fortunately, tackling this challenge probably does not need to design entirely new modelling paradigms but could build upon the current ones. One could devise an intermediate layer that interfaces the heterogeneous raw data and a network. This layer aims to amalgamate all available channels and modalities to form an input with a fixed number of channels which are ready to be fed to any existing network architecture. As a result, existing state-of-the-art models would be invigorated owing to the more informative input. Guillot and Thorey \cite{guillot_robustsleepnet_2021} proposed such a layer using a multi-head attention layer with $N$ heads to map an input with varying number of channels to $N$ channels. Potentially, this could also be done using an across-channel pooling operator, \emph{e.g.}, average pooling or max pooling. Channel mapping, for example via a convolutional  $1\times1$ kernel \cite{Lin2014}, could also map the varying inputs into any fixed number of channels and further enrich the resulting channels via non-linear activations. Ideally, such an interface layer should be integrated to an existing sleep staging model and trained jointly on the classification task in an end-to-end fashion. However, automatic channel quality control should be put in place to exclude bad-quality channels with, for example, highly noisy \cite{Stephansen2018} and excessive data missing \cite{Mikkelsen2021}.
	
	\subsection{Subjectivity in model building}
	\label{ssec:subjectivity}
	It is well-known that manual scoring of PSG is highly subjective. Many previous studies consistently reported the consensus among human scorers around a Cohen's kappa of $0.76$ \cite{DankerHopfe2009, Rosenberg2014}. The consensus is particularly poor on epochs manifesting characteristics of more than one sleep stages. Examples are N1 stage, epochs close to the boundary of two stages, and data from patients with fragmented sleep. So far, it has been a common practice to use the subjective and noisy labels annotated by a single scorer for model training as if they are perfect. Thus, the scorer's subjectivity is unavoidably transferred into the trained model. Reducing such subjectivity is necessary but remains largely uncharted.
	
	The above-mentioned subjectivity can be alleviated by scoring at a fine-grained temporal resolution (\emph{i.e.}, smaller epochs \cite{Stephansen2018, Olesen2021} or even samples \cite{Perslev2019,Perslev2021}). However, the fine-grained supervision signals in these works were still derived from 30-second epoch annotation, and therefore, the subjectivity continued to exist. A more proper treatment would be to consider the labels as noisy by nature (in practice, one-hot ground-truth labels are unknown or even not in existence) and robust training methods should be devised to manage the noisy label \cite{Nigam2020,Han2020}. It also remains an open question how to train a model with multiple supervision signals (\emph{i.e.}, annotations of two or more human scorers) at the same time rather than a single supervision signal as usual. Doing this will encourage the model to adapt to the scoring style of a cohort of scorers. Note that, this is different from averaging labels of multiple scorers \cite{Guillot2020} which eventually results in a single supervision signal.
	
	\subsection{Scorer personalization}
	Since it is very likely that a model trained with the annotation from a single scorer as described in Section \ref{ssec:subjectivity} will mimic (\emph{i.e.}, overfit) the scorer's style, the resulting ``subjective''  model poses another challenge from the end-user perspective that is worth being discussed separately here. Imagine a clinical scenario when a clinician is an end-user of the scoring system. The trained model can be reasonably viewed as a digital twin of the original scorer who labelled the data; hence, it will face disagreement on staging decisions with a new human scorer (\emph{i.e.}, an end-user in this case) as similar as disagreement between two human scorers. For maximum adoption by the clinician, this raises the need for the model to gradually adjust to the scoring style of the new scorer. Readers should  note that this scorer personalization problem is orthogonal to the data personalization discussed previously in Section \ref{sssec:individual_modelling}.
	
	Tackling this challenge requires a closed-loop interaction between the model and the end-user \cite{Liang2019}. On the one hand, the disagreed staging decisions need to be first identified. This could be done via uncertainty quantification, for example using entropy-based metrics \cite{Phan2021b,Mikkelsen2020} or an ensemble with Monte Carlo dropout \cite{Fiorillo2021}, as the model's decisions with low confidence are more likely to be disagreed ones. Although this approach can isolate a large portion of wrong decisions, a pitfall here is that an algorithm may output contentious decisions with a very high confidence. This anomalous behavior has been studied in many prior works \cite{Nguyen2015}. While understanding this behavior in the context of human sleep is an interesting subject on its own, methods will also need to be developed to identify wrong decisions associated with high confidence. Model's interpretability (see Section \ref{ssec:interpretability}) with the aid of convenient user-interface and user-experience design will be equally important to allow the end-user to scrutinize the potentially wrong decisions and make necessary corrections in an interactive manner. On the other hand, given the end-user's feedback, learning methods will be needed to incrementally adapt the model using the newly labelled data in an open-ended fashion. Approaches for continual learning \cite{Parisi2019}, such as the meta-learning method used in \cite{Banluesombatkul2021}, stands out as promising candidates. While these methods are required to learn from sequential, and potentially small, data, they also need to overcome catastrophic forgetting \cite{Kirkpatrick2017}, the central issue in this learning setting.

	\section{Conclusions}	
	
	Benefits of automatic tools for sleep scoring have driven research in automatic sleep staging for many decades. Promising results recently achieved by deep learning based methods have given incentives for a large number of research studies, now resulting in solutions that have comparable performance to sleep experts on the sleep scoring task, at least on healthy individuals. This methodological development has benefited immensely from the initiatives for free access to large collections of de-identified sleep data and open-source tools and techniques of many experiments available for researchers around the world.
	We perceive this achievement as a small yet important milestone in the use of AI in sleep research. However, many challenges still have to be overcome for these AI tools to prove their clinical usefulness. Next to the challenge of sleep disorders, issues related to data heterogeneity, model's explainability and subjectivity will require more attention. In order to bring sleep monitoring outside sleep labs to daily living environments, prospective studies also have to be conducted to improve robustness to low signal quality of mobile EEG devices and limited amounts of training data and to address issues related to privacy and longitudinal monitoring. Once these challenges have been overcome, we expect AI-based sleep scoring tools will play an important role in day-to-day sleep practice, complementing the increasing need for trained sleep experts and benefiting millions in need of accurate sleep assessment, monitoring, and treatment options for many sleep disorders. 
	
	\section*{Author contributions}
	H. Phan and K. Mikkelsen contributed equally. 
	
	\section*{Acknowledgement}
	H. Phan is supported by a Turing Fellowship under the EPSRC grant EP/N510129/1.
	
	\section*{References}
	\small
	\bibliographystyle{IEEEbib}
	\bibliography{bibliography}

\end{document}